\documentclass[11pt]{article}
\usepackage{jheppub}

\usepackage{epsfig}
\usepackage{graphicx}
\def\be{\begin{equation}}
\def\ee{\end{equation}}
\def\bea{\begin{array}}
\def\eea{\end{array}}
\def\beqa{\begin{eqnarray}}
\def\eeqa{\end{eqnarray}}
\def\beqas{\begin{eqnarray*}}
\def\eeqas{\end{eqnarray*}}

\def\bp{\begin{picture}}
\def\ep{\end{picture}}
\def\bc{\begin{center}}
\def\ec{\end{center}}
\def\bfig{\begin{figure}}
\def\efig{\end{figure}}

\def\bit{\begin{itemize}}
\def\eit{\end{itemize}}
\def\nn{\nonumber}
\def\f{\frac}

\def\[{\left[}
\def\]{\right]}
\def\({\left(}
\def\){\right)}

\def\..{\left.}
\def\.{\right.}
\def\tl{\tilde}
\def\ra{\rightarrow}

\def\NPB#1,{{ Nucl.\ Phys.\ B }{\bf #1},}
\def\PLB#1,{{ Phys.\ Lett.\ B }{\bf #1},}
\def\EPJC#1,{{ Eur.\ Phys.\ Jour.\ C }{\bf #1},}
\def\PRD#1,{{ Phys.\ Rev.\ D }{\bf #1},}
\def\PRL#1,{{ Phys.\ Rev.\ Lett.\ }{\bf #1},}
\def\MPLA#1,{{Mod.\ Phys.\ Lett.\ A }{\bf #1},}

\def\da{\dagger}

\def\al{\alpha}

\def\ep{\epsilon}

\def\pr{\prime}

\begin{document}

\title{Reconcile muon g-2 anomaly with LHC data in SUGRA with generalized gravity mediation}

\author[a,b]{Fei Wang,}
\emailAdd{feiwang@zzu.edu.cn}

\author[c]{Wenyu Wang,}
\emailAdd{wywang@mail.itp.ac.cn}

\author[b]{Jin Min Yang}
\emailAdd{jmyang@itp.ac.cn}

\affiliation[a]{Department of Physics and Engineering, Zhengzhou University,
Zhengzhou 450000, P. R. China}

\affiliation[b]{State Key Laboratory of Theoretical Physics, Institute of Theoretical Physics,
Chinese Academy of Sciences, Beijing 100190, P. R. China}

\affiliation[c]{Institute of Theoretical Physics, College of Applied Science,
Beijing University of Technology,
Beijing 100124, P. R. China}

\abstract{From generalized gravity mediation we build a SUGRA scenario in which the gluino
is much heavier than the electroweak gauginos at the GUT scale.
We find that such a non-universal gaugino scenario with very heavy gluino at the GUT scale
can be naturally obtained with proper high dimensional operators in the framework of SU(5) GUT.
Then, due to the effects of heavy gluino, at the weak scale all colored sparticles are heavy
while the uncolored spartilces
are light, which can explain the Brookhaven muon $g-2$ measurement while satisfying the
collider constraints (both the 125 GeV Higgs mass and the direct search limits of sparticles)
and dark matter requirements.
We also find that, in order to explain the muon $g-2$ measurement, the neutralino dark matter
is lighter than 200 GeV in our scenario, which can be mostly covered by the
future Xenon1T experiment. }

\maketitle

\section{Introduction}
 If the particle discovered by the ATLAS and CMS collaborations of the LHC \cite{atlas,cms}
is indeed the long missing standard model (SM) Higgs boson, then
the hierarchy problem related to the fundamental scalar may readily indicate
some new physics beyond the SM.
Another hint of new physics arises from the precise measurement of the
muon anomalous magnetic moment by the Brookhaven experiment \cite{brookhaven1,brookhaven2},
which gives a larger value than the SM prediction and the discrepancy is about $3\sigma$
\cite{smg-2}.

Among the new physics theories, the low energy supersymmetry (SUSY), which was initially
proposed to solve the gauge hierarchy problem, is still regarded as one of the most
appealing extensions.
The observed 125 GeV Higgs boson at the LHC  falls miraculously within the narrow
$115-135$ GeV "window" predicted by the minimal supersymmetric standard model (MSSM).
Besides, in the framework of low energy SUSY, the three gauge couplings
can naturally be unified \cite{su5,gutmssm,Einhorn:1981sx,Marciano:1981un},
the cosmic cold dark matter can be naturally explained, the vacuum instability
problem can be solved, and the muon $g-2$ discrepancy can also be accounted.

However, low energy SUSY also encounter some difficulties in the LHC era.
The heavy top-squarks needed by a 125 GeV Higgs boson\footnote{Note that
in the non-minimal SUSY models like the next-to-minimal SUSY model
such heavy top-squarks are not needed due to the additional tree-level
contributions to the Higgs mass (for a comparative study of different SUSY models
confronted with the LHC Higgs data, see, e.g., \cite{cao}).},
the null search results of sparticles and the perfect agreement of
$B_s^0\ra \mu^+\mu^-$ with the SM prediction all imply SUSY at a rather high scale.
Actually, the LHC data has already pushed
the gluino and squarks of first two generations to TeV scale \cite{cmssm1,cmssm2},
i.e., $m_{\tilde g} > 1.5$ TeV for $m_{\tl{q}} \sim m_{\tl{g}}$ and
$m_{\tl{g}}\gtrsim 1$ TeV for $m_{\tl{q}} \gg m_{\tl{g}}$, while for top-squarks the
bounds from the LHC search are model-dependent, e.g., above 600 GeV in
natural SUSY \cite{han}.
On the other hand, if the muon $g_\mu-2$ anomaly is solved in the framework of SUSY,
the relevent electroweak sparticles (smuons, neutralinos and charginos) need to
be around ${\cal O}(100)$ GeV for a $\tan\beta$ value of order ${\cal O}(10)$.
So it seems that low energy SUSY should be realized in a more involved way \cite{recent}.

If SUSY is required to account for the muon $g-2$ anomaly without contradiction
with the recent LHC results, a split spectrum for spartilces is favored, which
has one scale (relatively high) governing the colored sparticle masses and
the other scale (relatively low) governing the uncolored sparticle masses \cite{ewsusy}.
This can be realized in a supergravity (SUGRA) grand unified model
called gluino-SUGRA \cite{gsugra}
which
has non-universal gaugino masses\cite{ngaugino}, with the gluino being much heavier
than the electroweak gauginos at the GUT scale \cite{gsugra}.

In this note we try to build such a gluino-SUGRA model from the generalized gravity
mediation of SUSY breaking \cite{GSUGRA2}.
Oue results show that this scenario can be naturally obtained with proper
high dimensional operators in the framework of SU(5) GUT.
Then, due to the effects of heavy gluino, at the weak scale all colored sparticles are
heavy while the uncolored spartilces
are light, which can explain the Brookhaven muon $g-2$ measurement while satisfying the
collider constraints and dark matter requirements.
We also find that, in order to explain the muon $g-2$ measurement, the neutralino dark matter
is below 200 GeV in this scenario, which can be mostly covered by the
future Xenon1T experiment.

 This paper is organized as follows.  In Sec.\ref{typeII}, we construct
a gluino-SUGRA model in the framework of SU(5) GUT from the generalized gravity mediation.
 In Sec.\ref{dm}, we examine the phenomenological constraints on our scenario,
 which are from the muon $g-2$, the LHC data, and the dark matter relic density
and direct detection limits.
Sec.\ref{conclusion} contains our conclusions.

\section{SUGRA with heavy-gluino constructed from generalized gravity mediation}
\label{typeII}
To mediate the SUSY breaking effects from the hidden sector to the visible sector,
many types of mechanisms are proposed, for example,
gravity mediation \cite{mSUGRA}, gauge mediation \cite{gaugemediation}
and anomaly mediation \cite{anomalymediation}. Among these  mechanisms,
the gravity mediation is a very predictive scenario.
In this scenario the SM-like Higgs boson mass lies close to the upper limit 130 GeV
predicted in grand unified SUGRA models \cite{higgssugra}.

In the popular gravity mediation scenario, the K\"ahler potential is assumed
to be minimal. When certain high-representation chiral
fields of the GUT group are involved in the nonrenormalizable K\"ahler potential,
the kinetic terms of superfields can have alternative contributions
after the GUT symmetry breaking. New nonrenormalizable terms in the
superpotential involving high-representation fields can also be important.
In general, both gauge singlet and non-singlet can acquire non-vanishing F-term
VEVs to break supersymmetry. We will focus on the SU(5) grand unified SUGRA model
in our analysis.

A general form of the kinetic terms for vector supermulitplet is
\beqa
\label{gaugekinetic}
{\cal L}\supseteq \int d^2\theta \tau \( W^aW^a+a_1\f{S}{M_*}W^aW^a+b_1 \f{\Phi_{ab}}{M_*}W^a W^b\),
 \eeqa
with $'\Phi'$ denoting a GUT group non-singlet chiral supermultiplet and $'S'$ a GUT group singlet which can acquire a VEV of order (or below) $M_*$.

From the symmetric product of SU(5) adjoint
 \beqa
\( {\bf 24} \otimes {\bf 24}\)_{\bf symmetric}={\bf 1}\oplus {\bf 24}\oplus {\bf 75}\oplus{\bf 200}~,
 \eeqa
we can see that the non-renormalizable terms can be constructed with {\bf 24,75,200}
representation chiral supermultiplets of SU(5).
For simplicity, we assume that only ${\bf 75}$ representation chiral field appears
in Eq.(\ref{gaugekinetic}) and in the Kahler potential of the form
\beqa
\label{kinetic}
K=\phi^\da \phi+\f{c_1}{M_{*}}\sum\limits_{r} \phi_a^\da S \phi_a+\f{c_1^\pr}{M_*^2}\sum\limits_{r} \phi_a^\da (\overline{\Phi}^{\bf 75}\otimes \Phi^{\bf 75})^r_{ab} \phi_b ,
\eeqa
with $r$ denoting some representation from production expansion of ${\bf 75}\otimes {\bf 75}$. We assume that the superfield $\Phi_{\bf 75}$ acquires both the lowest component and F-term VEVs. After the GUT singlet $S$ field and ${\bf 75}$ field acquire the lowest component VEVs
\beqa
<\Phi_{\bf 75}>_{ab}=v_{\bf 75}U_{ab}~,
\eeqa
with the universal group factor $U_{ab}$ given in terms of $10\times 10$ matrix as
\beqa
U_{ab}=\f{1}{\sqrt{12}}\(~1,~1,~1,-1,-1,-1,-1,-1,-1,~3\)~,
\eeqa
 the wave-function normalization factor for the gauge kinetic term will have the form
\beqa
Z_i=1+a_1\f{<S>}{M_*}+b_1\f{<\Phi>_i}{M_*}\equiv \al+\beta_i,
\eeqa
with the ratios of the non-universal parts given by $\beta_1:\beta_2:\beta_3=-5:3:1$.

 The F-term VEV of $\Phi^{\bf 75}$ given by $(F_{\Phi})_{ab}=F_{\Phi}\cdot U_{ab}$ will
lead to a non-canonical gaugino mass ratio
 \beqa
 M_1:M_2:M_3=-b_1\f{5}{4\sqrt{3}}\f{F_{\bf 75}}{M_*}:b_1\f{3}{4\sqrt{3}}\f{F_{\bf 75}}{M_*}:b_1\f{1}{4\sqrt{3}}\f{F_{\bf 75}}{M_*}~.
 \eeqa
 which, after re-scaling the normalization factor, will give a physical non-universal
gaugino mass ratio
 \beqa
 M_1:M_2:M_3=\f{-5}{Z_1}:\f{3}{Z_2}:\f{1}{Z_3}.
 \eeqa
 For the choice of coefficient $Z_3\equiv\al+\beta_3\approx {\cal O}(0.1)\approx 0$, we can fix the value of $\al$ and thus the value of $Z_1,Z_2$ correspondingly.  The ratio for $Z_i$ will be given approximately by
 \beqa
 Z_1:Z_2:Z_3\approx -6: 2: Z_3,
 \eeqa
so we can obtain
\beqa
M_1: M_2: M_3=\f{5}{6}:\f{3}{2 }:\f{1}{Z_3}.
\eeqa
We can see that at the GUT scale the gluino can be much heavier than bino and wino. On the other hand, the gluino will in general not be too heavier than the other two if no fine-tuning in the normalization factor is introduced.

From group theory we know
\beqa
{\bf 75}\otimes {\bf 75} \supset {\bf 1\oplus 24\oplus 75\oplus 200},
\eeqa
so a unnormalized universal sfermion mass can be generated from Kahler potential by F-term VEVs
of ${\bf 75}$:
\beqa
{m}^2_{\tl{10}_i,\tl{ \overline{5}}_i,\tl{H}_{u,d}}&=&\f{3}{2}c_{\bf 1}^\pr\f{|F_{\bf 75}|^2}{M_*^2}.
\eeqa
 Note that there are many possible contractions of group factors in the Kahler potential
and we adopt here the simplest case with $(\overline\Phi_{\bf 75}\otimes \Phi_{\bf 75})_{ab}\propto \delta_{ab}$. On the other hand, it can be seen from formula (\ref{kinetic}) that the kinetic terms for matter contents will also get additional contributions from the GUT breaking effects by the lowest component VEVs of {\bf 75}. So the unnormalized universal sfermion masses should be rescaled with respect to the kinetic factor to get the physical soft masses
\beqa
{m}^2_{\tl{10}_i,\tl{ \overline{5}}_i,\tl{H}_{u,d}}&=&\f{3}{2 Z_\phi }c_1^\pr\f{|F_{\bf 75}|^2}{M_*^2}.
\eeqa
with possible kinetic factor $Z_\phi$ as
\beqa
Z_\phi=1+\f{c_1 \langle S\rangle}{M_*}+c^\pr_1\f{ v_{\bf 75}^2}{M_*^2}\approx 1+\f{c_1  \langle S\rangle}{M_*}\equiv Z_U.
\eeqa
Therefore, the universal sfermion mass can be set as an free parameter in our scenario.
The universal sfermion masses, which control the masses for slepton, should not be heavy in order to explain the $g_\mu-2$ anomaly. The squarks, on the other hand, will receive large corrections from gluino loops.
So the typical universal sfermion mass scale should not be too larger than that of the lightest gaugino.

 The trilinear term can be generated from the non-renormalization operators in the
superpotential involving {\bf 75} superfield
  \beqa
  W\supset \(\f{\Phi^{\bf 75}_{ab}}{M_*}+c_1 \f{S\Phi^{\bf 75}_{ab}}{M_*^2}\)\sum\limits_{i,j=1}^3\(y_{ij}{\bf 10}_i\otimes{\bf 10}_j\otimes{\bf 5}_{H_u} +y_{ij}^\pr {\bf 10}_i\otimes\overline{\bf 5}_j\otimes\overline{\bf 5}_{H_d}\)_{ab} ,
  \eeqa
with $i,j$ denoting the family index. Relevant calculations can be found in our previous works\cite{GSUGRA2}. Similar calculations give the resulting trilinear terms
\beqa
-{\cal L}&\supset& \f{F_{75}}{M_*(Z_U)^{3/2}}(1+c_1 \f{\langle S\rangle}{M_*})\( 3 y_{ij}^{E}\tl{L}_i\tl{E}_j H_d-y_{ij}^D \tl{Q}^i\tl{D}_j H_d\)\nn\\
&\approx& \f{F_{75}}{M_*(Z_U)^{1/2}}\( 3 y_{ij}^{E}\tl{L}_i\tl{E}_j H_d-y_{ij}^D \tl{Q}^i\tl{D}_j H_d\).
\eeqa
after rescaling the kinetic factor $Z_U$ for sfermions and higgs chiral fields ${\bf 5}_{H_u},{\bf \overline{5}}_{H_d}$.  Note that our previous calculations \cite{GSUGRA2} indicate that the up-type squark trilinear terms vanish if we only introduce the F-term VEV for {\bf 75}.

The SUSY preserving $\mu$ term, which will be determined by the electroweak symmetry
breaking conditions, is generated by fine tuning with the lowest component VEV
of $\Phi^{\bf 24}$
\beqa
W\supset (M+\langle\Phi^{\bf 24}\rangle){\bf 5}_{H_u} \overline{\bf 5}_{H_d} .
\eeqa
Because one cannot construct gauge invariant combinations involving only ${\bf 5,\bar{5}}$
and ${\bf 75}$, the $B_\mu$ term can be generated from
\beqa
W\supset \f{1}{M_*^2}(M+\langle\Phi^{\bf 24}\rangle)\Phi^{\bf 24}\Phi^{\bf 75}{\bf 5}_{H_u} \overline{\bf 5}_{H_d},
\eeqa
which gives
\beqa
B_\mu=\f{5}{6\sqrt{3}}\f{v_{24}}{M_*}\f{F_{\bf 75}}{M_*}\mu.
\eeqa
So we can see that the $B_0\equiv B_\mu/\mu$ term at the GUT scale is suppressed by a
GUT/Planck factor relative to $A_0$ and can be set to zero at the GUT scale.

We can introduce only the ${\bf 24}$ or ${\bf 200}$ representation field as
the GUT non-singlet field $\Phi$ in the generalized gauge kinetic terms
and then the GUT scale non-universal gaugino \cite{ngaugino1,ngaugino2} input
will be changed accordingly:
\bit

\item  The scenario with only  ${\bf 24}$ representation Higgs:\\
 The lowest component VEV for the ${\bf 24}$ representation field has the form
\beqa
<\Phi_{\bf 24}>_{ab}=v_{\bf 24}U_{ab}~,
\eeqa
with the universal group factor $U_{ab}$ given in terms of $5\times 5$ matrix by
\beqa
U_{ab}=\f{1}{\sqrt{15}}\(~1,~1,~1,-\f{3}{2},-\f{3}{2}\)~.
\eeqa
Similar to the case of ${\bf 75}$ representation Higgs, the ratios of the
non-universal parts within the wave-function normalization factor of
gauge kinetic terms will be given by $\beta_1:\beta_2:\beta_3=1:3:-2$.
This will lead to GUT scale non-universal gaugino input:
\beqa
M_1:M_2:M_3=\f{1}{3}:\f{3}{5}:\f{-2}{Z_3}\simeq{\cal O}(10).
\eeqa
\item The scenario with only  ${\bf 200}$ representation Higgs:\\
 The lowest component VEV for the ${\bf 200}$ representation field has the form
\beqa
<\Phi_{\bf 200}>_{ab}=v_{\bf 200}U_{ab}~,
\eeqa
with the universal group factor $U_{ab}$ given in terms of $15\times 15$ matrix by
\beqa
U_{ab}=\f{1}{\sqrt{12}}\(\underbrace{~1,\cdots,~1}_6,\underbrace{-2,\cdots,-2}_6,\underbrace{2,\cdots, 2}_3\)~.
\eeqa
The ratios of the non-universal parts within the wave-function normalization factor of
gauge kinetic terms are given by
$\beta_1:\beta_2:\beta_3=10:2:1$. This will lead to a GUT scale non-universal gaugino input:
\beqa
M_1:M_2:M_3=\f{10}{9}:\f{2}{1}:\f{1}{Z_3}\simeq{\cal O}(10).
\eeqa
\eit
So we see that such a non-universal gaugino scenario with very heavy gluino at the GUT scale
can be naturally obtained with proper high dimensional operators in the framework of SU(5)
GUT.

\section{Phenomenological constraints}\label{dm}
Now we scan the parameter space of our gluino-SUGRA scenario.
The GUT scale inputs can are given by
\bit
\item The gaugino mass scale $M_{1/2}$ with non-universal gaugino mass raito
\beqa
M_1:M_2:M_3=\f{5}{6}:\f{3}{2}:\f{1}{Z_3}(\sim {\cal O}(10)) ,
\eeqa
for the case with the ${\bf 75}$ representation Higgs.
Here we define  $M_{1/2}$ as $M_1=(5/6) M_{1/2}$ and in our numerical calculations
we will vary $1/Z_3$ from 10 to 50.
At the weak scale, the gaugino mass ratio is estimated to be
\beqa
M_1:M_2:M_3\approx \f{5}{6}:3:\f{6}{Z_3}.
\eeqa
\item The universal sfermion mass $M_S$.
\item The trilinear term $A_{b,\tau}$ (at the same order as $M_S$) while $A_t=0$.
\item The $B_0$ parameter is set to zero at the GUT scale.
\item The parameter $\tan\beta$ (in our scan we vary it in the range $1<\tan\beta<50$). 
 Choices of $\tan\beta$ which can not trigger successful radiative EWSB will not be kept in our numerical scan.
\eit
The $\mu$ parameter is determined by the electroweak symmetry breaking conditions.

We use the code DarkSUSY \cite{darksusy} to scan over the parameter space
and use the code SuSpect2\cite{suspect} to obtain the low energy spectrum
by RGE running from the GUT scale (at this energy scale $g_1 = g_2$ ) to the weak scale.
The central values of $g_1,g_2$ and $g_3$ at the weak scale are used as the inputs.
Other inputs, for example, the top Yukawa coupling $h_t$, are extracted
from the standard model taking into account the threshold corrections
(the relevant details can be seen in the appendix of \cite{FWY,bernal}).

In our scan we consider the following constraints (the relevant details can be found
in our previous work \cite{FWY1}):
\bit
\item[(1)] The relic density of the neutralino dark matter given by
           Planck $\Omega_{DM} = 0.1199\pm 0.0027$ \cite{planck}
           (in combination with the WMAP data \cite{wmap}).
\item[(2)] The LEP lower bounds on neutralinos and charginos ( $m_{\chi^C}> 103 {\rm GeV}$)
           as well as the bounds from invisible $Z$ decay
          $\Gamma(Z\ra \chi^0\chi^0)<1.71~{\rm MeV}$  which is consistent with
          the $2\sigma$ precision EW measurement result $\Gamma^{non-SM}_{inv}< 2.0~{\rm MeV}$.
\item[(3)] The precison electroweak observables $S,T,U$  \cite{oblique} to be compatible
           with the LEP/SLD data at 2$\sigma$ level \cite{stuconstraints}.
\item[(4)] The LHC constraints on the SM-like Higgs boson mass
           $123 {\rm GeV}<M_h <127 {\rm GeV}$ \cite{atlas,cms}.
\eit
In our scan, we also require that the survived points should satisfy successful EW symmetry breaking requirements which otherwise will not be kept.
Under the above constraints we will show the SUSY contributions to the muon $g_\mu-2$
and  the spin-independent dark matter-nucleon scattering rates compared with
the dark matter direct detection limits from Xenon100 \cite{xenon} and
the LUX \cite{lux}:
\bit
\item For the spin-independent dark matter-nucleon scattering
rate, we calculate it with the parameters \cite{Djoudi,Carena,Hisano:2010ct}:
$f_{T_u}^{(p)} =0.023,f_{T_d}^{(p)} = 0.032$,$f_{T_u}^{(n)} = 0.017, f_{T_d}^{(n)} = 0.041$
 and $f_{T_s}^{(p)} = f_{T_s}^{(n)} = 0.020$. The value of  $f_{T_s}$
is taken from the lattice simulation results \cite{lattice}.
 All the contributions known so far, including QCD corrections, are taken into account
in our calculation of the scattering rate.

\item For the SUSY contributions to the muon $g_\mu-2$, we know that they are
dominated by the chargino-sneutrino and neutralino-smuon loops.
At the leading order of $m_W/m_{SUSY}$ and $\tan\beta$ ($m_{SUSY}$
denotes the SUSY-breaking masses), the SUSY loop contributions are \cite{moroi,endo}
 \beqa
 \label{moroi}
 \Delta a_\mu(\tl{W},\tl{H},\tl{\nu}_\mu)&\simeq& 15\times 10^{-9}\left(\f{\tan\beta}{10}\right)\left(\f{(100 {\rm GeV})^2}{\mu~M_2}\right),~\\
 \Delta a_\mu(\tl{W},\tl{H},\tl{\mu}_L)&\simeq& -2.5\times 10^{-9}\left(\f{\tan\beta}{10}\right)\left(\f{(100 {\rm GeV})^2}{\mu~ M_2}\right),~\\
 \Delta a_\mu(\tl{B},\tl{H},\tl{\mu}_L)&\simeq& 0.76\times 10^{-9}\left(\f{\tan\beta}{10}\right)\left(\f{(100 {\rm GeV})^2}{\mu~ M_1}\right),~\\
 \Delta a_\mu(\tl{B},\tl{H},\tl{\mu}_R)&\simeq& -1.5\times 10^{-9}\left(\f{\tan\beta}{10}\right)\left(\f{(100 {\rm GeV})^2}{\mu~M_1}\right),~\\
 \label{moroi2}
 \Delta a_\mu(\tl{\mu}_L,\tl{\mu}_R,\tl{B})&\simeq& 1.5\times 10^{-9}\left(\f{\tan\beta}{10}\right)\left(\f{(100 {\rm GeV})^2(\mu~M_1)}{m_{\tl{\mu}_L}^2m_{\tl{\mu}_R}^2}\right),~
 \eeqa
 The SUSY contributions to the muon $g_\mu-2$ will be enhanced for small soft masses
and large $\tan\beta$.
Since the experimental value is larger than the SM prediction, a positive
$\mu M_{1,2}$ is favored in most of the parameter space. See also the results from the numerical code\cite{Farinaldo}.
\eit

%%%%%%%%%%%%%% Fig.1 %%%%%%%%%%%%%%%%%%%%%%%%%%%%%%%%%%%%%%%%
\begin{figure}[htbp]
  \begin{minipage}[t]{0.5\linewidth}
    \centering
    \includegraphics[width=2.8 in]{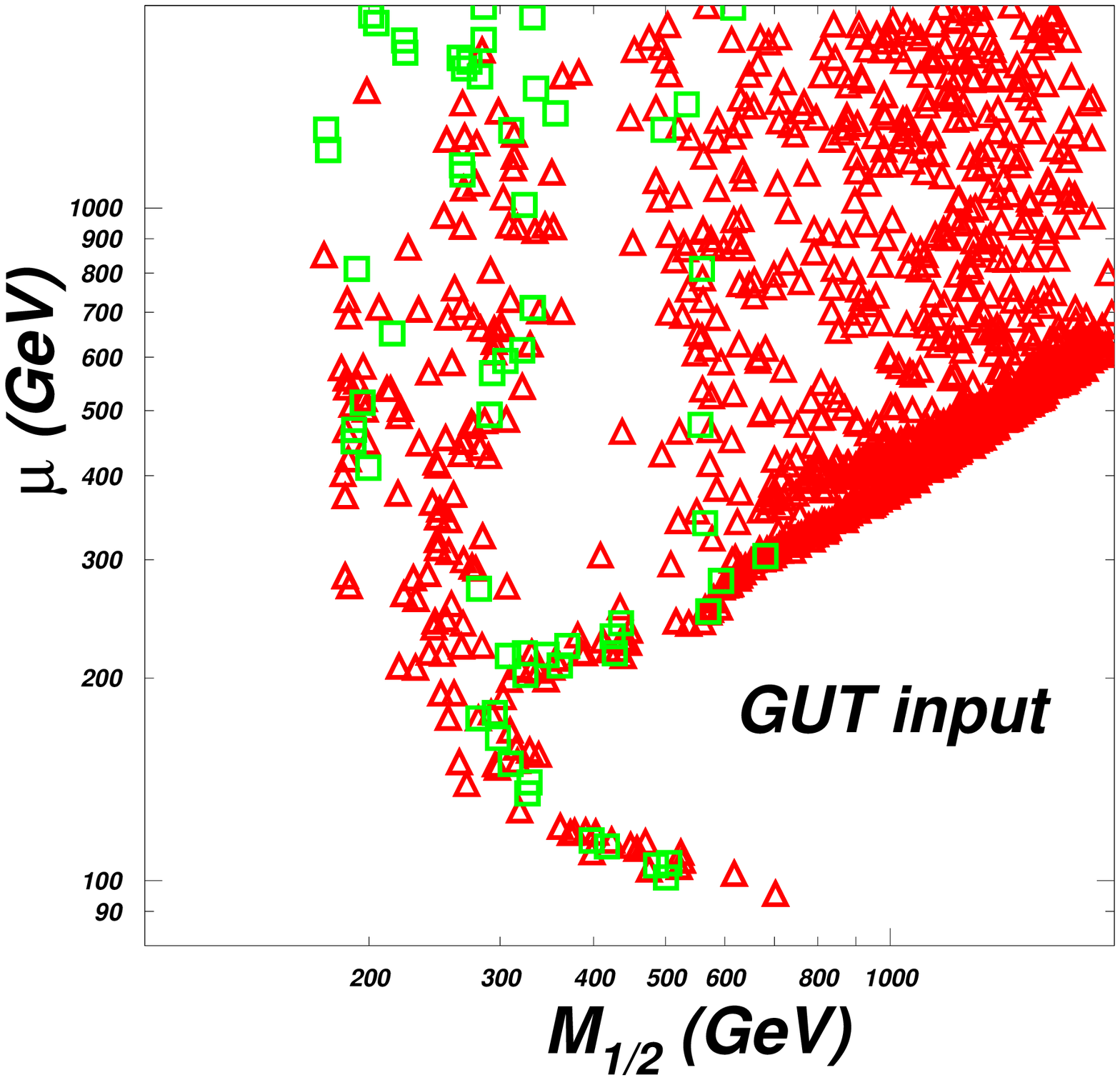}
     %%\label{fig:side:a}
  \end{minipage}
  \begin{minipage}[t]{0.5\linewidth}
    \centering
    \includegraphics[width=2.7 in]{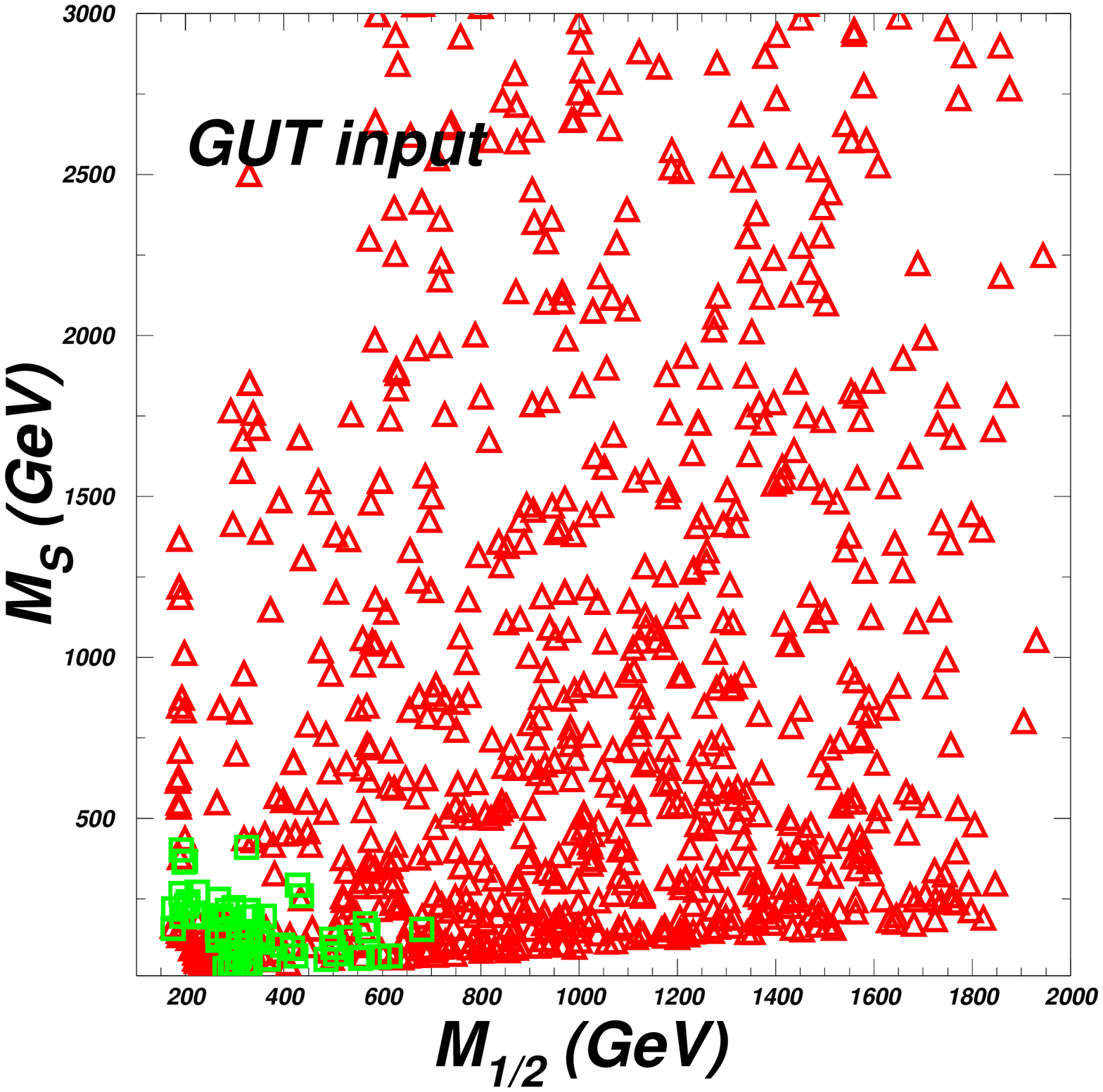}
    %%%\caption{aaaa}  %%   \label{fig:side:b}
  \end{minipage}
  \begin{minipage}[t]{0.5\linewidth}
    \centering
    \includegraphics[width=2.8in]{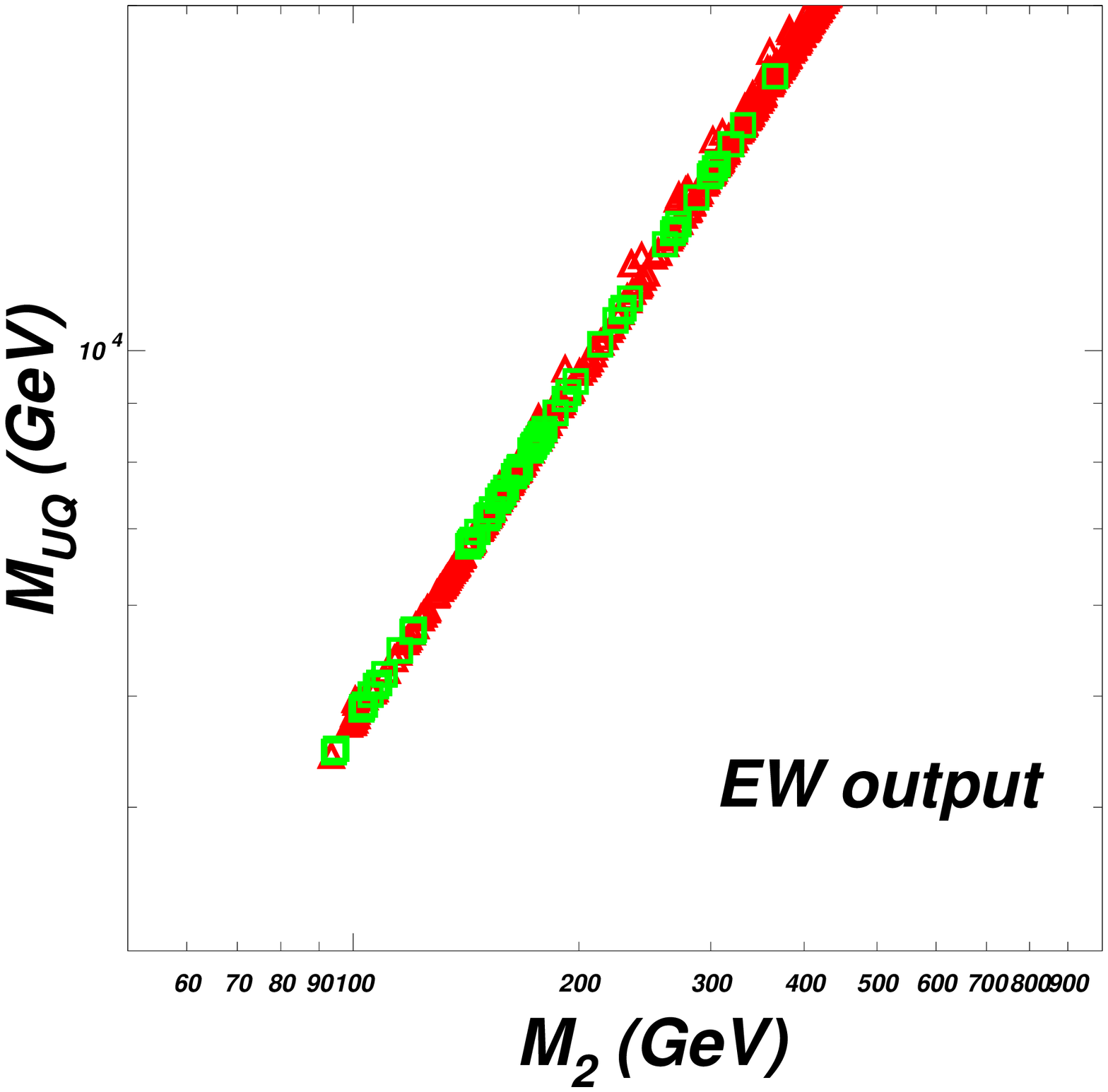}
    %%%\caption{aaaa}  %%   \label{fig:side:b}
  \end{minipage}
    \begin{minipage}[t]{0.5\linewidth}
    \centering
    \includegraphics[width=2.7in]{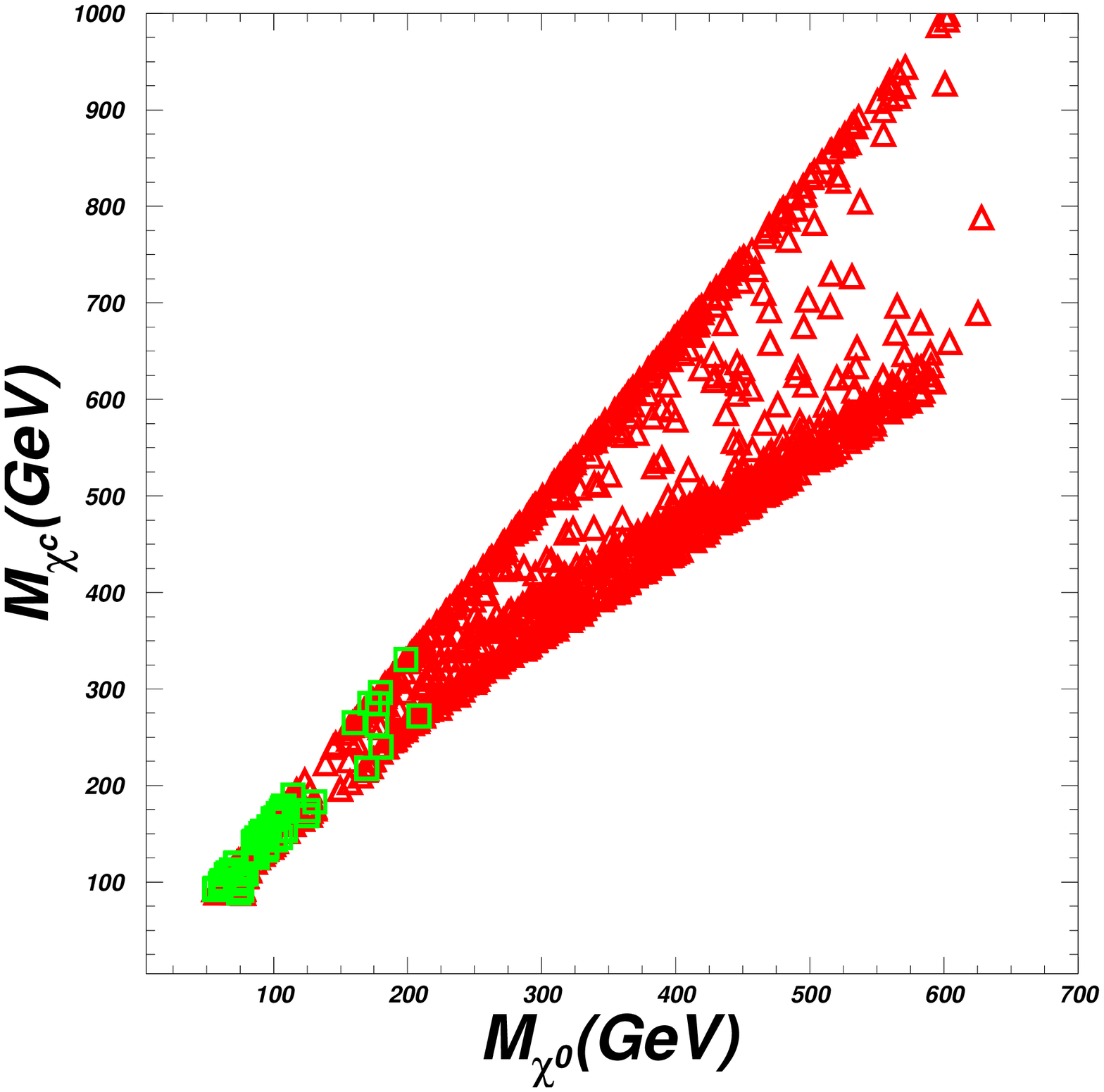}
    %%%\caption{aaaa}  %%   \label{fig:side:b}
  \end{minipage}
\caption{The scatter plots of the samples that satisfy the constraints (1-4) for $k\equiv 1/Z_3=10$.
The green $'\square'$  (red $'\triangle'$) can (cannot) explain the muon $g_\mu-2$
at $1\sigma$ level.
The upper panels show the input parameters at GUT scale
while the lower panels show the output parameters at electroweak (EW) scale, with
$M_{UQ}$ denoting the up-squark soft mass, $M_{\chi^C}$ the lightest chargino mass and
$M_{\chi^0}$ the lightest neutralino mass.
}
\label{fig1}
\end{figure}
%%%%%%%%%%%%%% Fig.2 %%%%%%%%%%%%%%%%%%%%%%%%%%%%%%%%%%%%%%%%
\begin{figure}[htbp]
 \centering
    \includegraphics[width=5.5in]{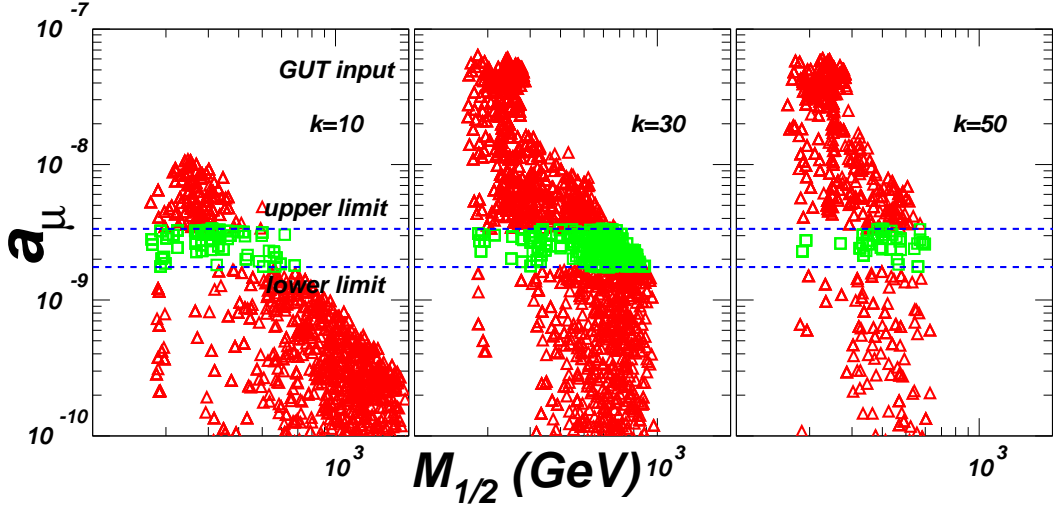}
\vspace*{-.5cm}
\caption{Same as Fig.1, but showing the muon $g_\mu-2$ versus $M_{1/2}$ for
different values of $k\equiv 1/Z_3$. The region between the two horizontal dashed lines
corresponds to the Brookhaven measured $g_\mu-2$ at $1\sigma$ level.}
\label{fig2}
\end{figure}
%%%%%%%%%%%%%% Fig.3 %%%%%%%%%%%%%%%%%%%%%%%%%%%%%%%%%%%%%%%%
\begin{figure}[htbp]
 \centering
    \includegraphics[width=5.5in]{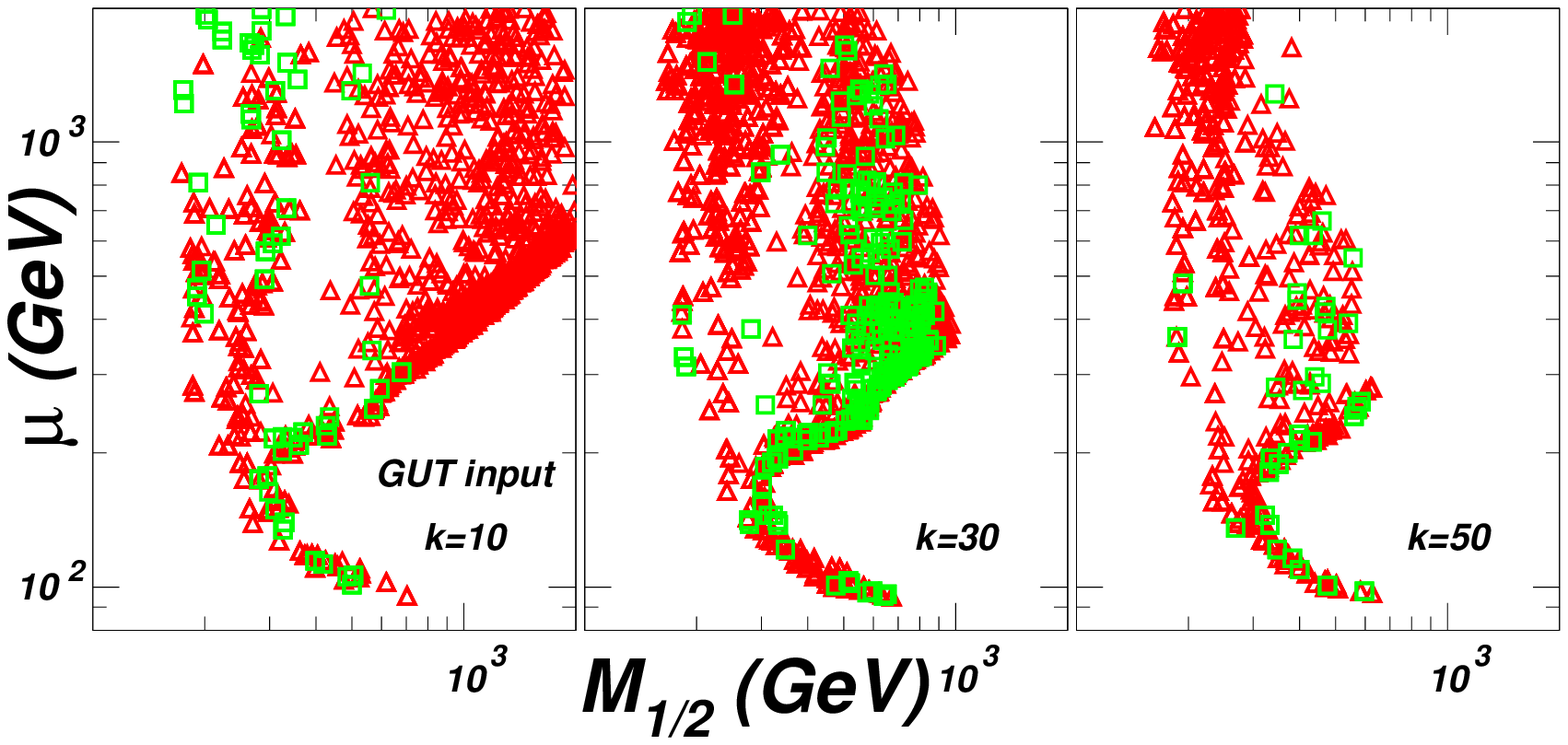}
    \includegraphics[width=5.5in]{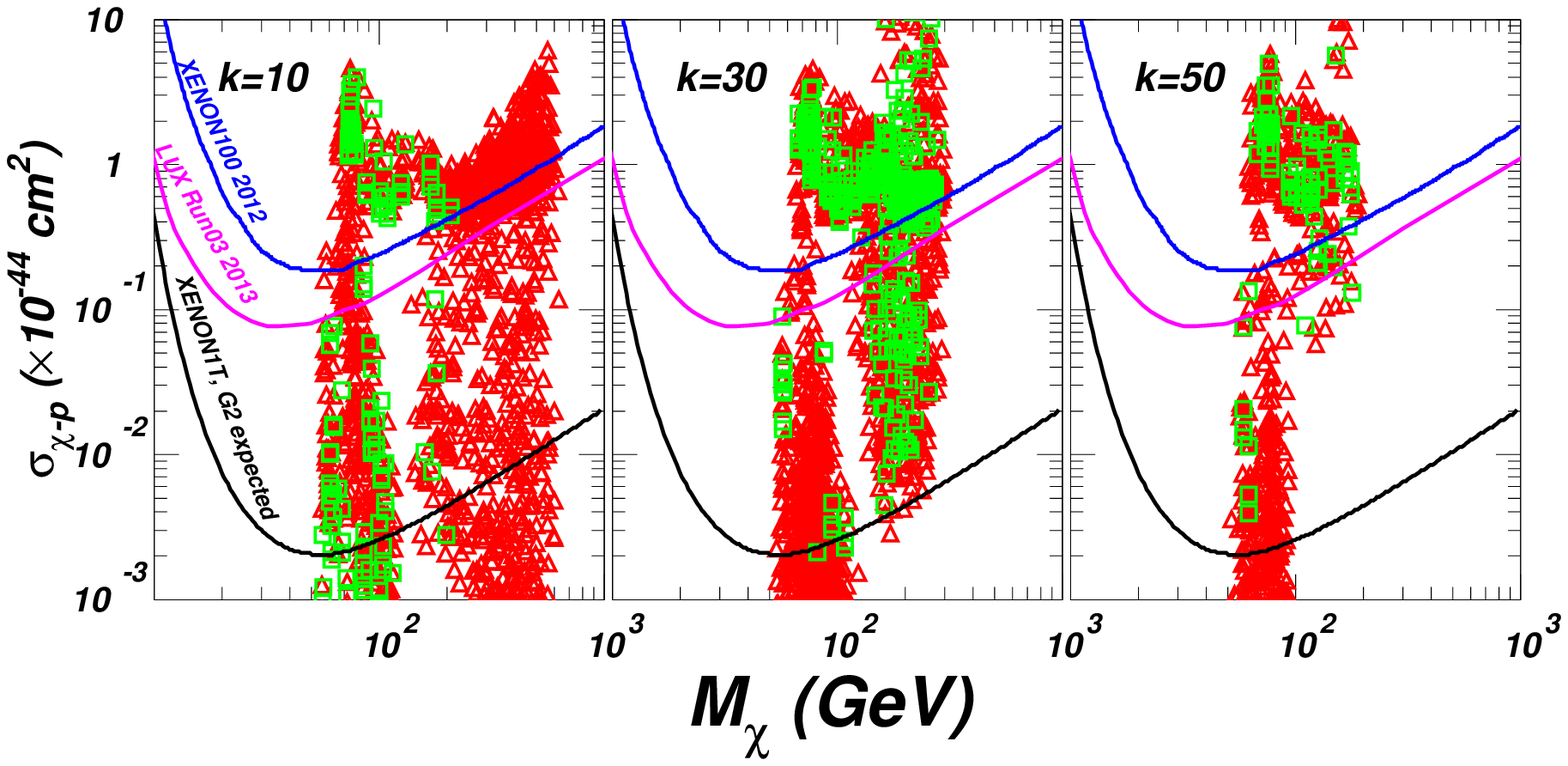}
\vspace*{-.5cm}
\caption{Same as  Fig.2, but showing $\mu$ versus $M_{1/2}$ in the upper panel
and the spin-independent dark matter-nucleon scattering rate versus the
lightest neutralino dark matter mass in the lower panel. The curves in the
lower panel show the dark matter direct detection limits from Xenon \cite{xenon}
and the LUX \cite{lux}.}
\label{fig3}
\end{figure}
%%%%%%%%%%%%%%%%%%%%%%%%%%%%%%%%%%%%%%%%%%%%%%%%%%%%%%%%%%%%%

The numerical results from our scan are shown in
Fig.\ref{fig1}, Fig.\ref{fig2}, Fig.\ref{fig3} and Table 1.
All the points in the
figures can satisfy the constraints (1-4), where the green $'\square'$
(red $'\triangle'$) can (cannot)  explain the muon $g_\mu-2$ deviation
 $\Delta a_\mu = (255\pm 80)\times 10^{-11}$ at $1\sigma$ level.
As shown in these figures, some samples
in our gluino-SUGRA scenario can
satisfy the LHC constraints and explain the Brookhaven $g_\mu-2$ experiment.
For these samples, we have the following observations:
\bit
\item[(i)]
From the upper left panel of Fig.1, we can see that the muon  $g_\mu-2$ explanation
constrains $M_{1/2}$ (defined as $M_1=(5/6) M_{1/2}$ at the GUT scale) below 600 GeV.
The upper right panel shows  the muon  $g_\mu-2$ explanation also requires
a light value for the universal sfermion mass $M_S$ at the GUT scale
(so that at the electroweak scale we have
light sleptons and electroweakinos while squarks are heavy due to RGE running).
Such results can be easily understood from Eqs.(\ref{moroi}-\ref{moroi2}).

\item[(ii)] Due to a rather heavy gluino, when squark masses run down to
 the electroweak scale, they become sufficiently heavy (although  $M_S$ is light
at the GUT scale) as required by a 125 GeV SM-like Higgs boson and the LHC bounds.
We can see from the lower left panel of  Fig.\ref{fig1} that the squarks at the electroweak
scale are heavier than 4 TeV in our scenario.

\item[(iii)] From  the lower right panel of Fig.\ref{fig1} we see that
 the muon  $g_\mu-2$ explanation requires low masses for the lightest chargino and
the lightest neutralino. The lightest neutralino dark matter lies in the mass range
of 80 to 200 GeV.

\item[(iv)] From Fig.\ref{fig3} we see that for $k=10$ ($k\ge 30$)
most (all) samples required to explain the muon $g_\mu-2$ at $1\sigma$ level
can be covered by the future Xenon1T experiment. This means that in case of
null results at Xenon1T experiment, our scenario with  $k\ge 30$ will be
excluded.

\item[(v)] The low energy particle spectrum for some typical benchmark points are shown in Table 1.  
We can see that the sleptons are typically light while the squarks are heavy due to the 
much heavier gluino.

\eit
\begin{table}
\label{benchmark}
  \centering
  \caption{The masses (in GeV) of some sparticles at the weak scale for different values of 
$k\equiv 1/Z_3$. All the points satisfy $g_\mu-2$ constraints and other
electroweak constraints.}
  \begin{tabular}{ccccccccccc}
  \hline\hline
 $k$ & $M_{1/2}$ & $\mu$  & $m_{\chi^0}$&$m_{\tilde g}$&  $m_{\tilde \mu_1}$
    & $m_{\tilde \mu_2}$& $m_{\tilde u_1}$ & $m_{\tilde d_1}$\\
\hline
10 & 297.63 & 163.09  &  79.54 &  8583.95 &  273.28 &    246.47 &  7594.43 & 7594.35  \\ \hline
30 & 535.07 & 367.18  & 190.49 & 46289.43 &   195.09 &    286.52 & 40934.25 & 40934.15 \\ \hline
50 & 398.92 & 213.81  & 132.09 & 57522.28 &   171.78 &    232.77 & 50867.91 & 50867.88 \\
\hline\hline
  \end{tabular}
\end{table}
\section{\label{conclusion}Conclusion}
From generalized gravity mediation we constructed a SUGRA scenario in which the gluino
is much heavier than the electroweak gauginos at the GUT scale.
We chose the framework of SU(5) GUT and found that such a non-universal gaugino
scenario with very heavy gluino at the GUT scale
can be naturally obtained with proper high dimensional operators.
Due to the contributions of the heavy gluino,
at the weak scale the squarks are sufficiently heavy as required by a 125 GeV
SM-like Higgs boson, while the uncolored spartilces can be light enough to
explain the Brookhaven muon $g-2$ measurement.
Since the muon $g-2$ explanation requires a neutralino dark matter
below 200 GeV in our scenario, the parameter space can be mostly covered by
the future Xenon1T experiment.

\section*{Acknowledgement}
This work was supported by the
Natural Science Foundation of China under grant numbers 11105124, 11105125,
11275245, 10821504,  11135003, 11375001, 11172008, by Ri-Xin Foundation of BJUT and
by tanlents foundation of eduction department of Beijing.


\begin{thebibliography}{99}
\vspace{-1mm}

\bibitem{atlas} G. Aad et al.(ATLAS Collaboration), Phys. Lett. B710, 49 (2012).

\bibitem{cms} S. Chatrachyan et al.(CMS Collaboration), Phys. Lett.B710, 26 (2012).

\bibitem{brookhaven1} Muon G-2 collaboration, G. Bennett et al., Phys. Rev. D 73 (2006) 072003.

\bibitem{brookhaven2} B.L. Roberts, Chin. Phys. C 34 (2010) 741.

\bibitem{smg-2} K. Hagiwara, A. Martin, D. Nomura and T. Teubner, Phys. Lett. B 649 (2007) 173;
T. Teubner, K. Hagiwara, R. Liao, A. Martin and D. Nomura, Chin. Phys. C 34 (2010) 728;
 M. Davier, A. Hoecker, G. Lopez Castro, B. Malaescu, X. Mo et al., Eur. Phys. J. C 66 (2010) 127;
 M. Davier, A. Hoecker, B. Malaescu, C. Yuan and Z. Zhang, Eur. Phys. J. C 66 (2010) 1.

\bibitem{su5}
H.~Georgi and S.~L.~Glashow, Phys.\ Rev.\ Lett.\ {\bf 32}, 438 (1974).
%%CITATION = PRLTA,32,438;%%

 \bibitem{gutmssm}
J.~R.~Ellis, S.~Kelley and D.~V.~Nanopoulos, Phys.\ Lett.\ B {\bf 249}, 441 (1990);
%%CITATION = PHLTA,B249,441;%%
Phys.\ Lett.\ B {\bf 260}, 131 (1991);
%%CITATION = PHLTA,B260,131;%%
U.~Amaldi, W.~de Boer and H.~Furstenau, Phys.\ Lett.\ B {\bf 260}, 447 (1991);
%%CITATION = PHLTA,B260,447;%%
P.~Langacker and M.~X.~Luo, Phys.\ Rev.\ D {\bf 44}, 817 (1991).
%%CITATION = PHRVA,D44,817;%%

\bibitem{Einhorn:1981sx}
  M.~B.~Einhorn and D.~R.~T.~Jones,
  %``The Weak Mixing Angle and Unification Mass in Supersymmetric SU(5),''
  Nucl.\ Phys.\ B {\bf 196} (1982) 475.
  %%CITATION = NUPHA,B196,475;%%

\bibitem{Marciano:1981un}
  W.~J.~Marciano and G.~Senjanovic,
  %``Predictions of Supersymmetric Grand Unified Theories,''
  Phys.\ Rev.\ D {\bf 25} (1982) 3092.
  %%CITATION = PHRVA,D25,3092;%%

\bibitem{cao}
  J.~Cao  {\it et al.},
  %``A SM-like Higgs near 125 GeV in low energy SUSY: a comparative study for MSSM and NMSSM,''
  JHEP {\bf 1203}, 086 (2012) [arXiv:1202.5821 [hep-ph]];
  %%CITATION = ARXIV:1202.5821;%%
%``Status of low energy SUSY models confronted with the LHC 125 GeV Higgs data,''
 JHEP {\bf 1210}, 079 (2012)  [arXiv:1207.3698 [hep-ph]].
%%CITATION = ARXIV:1207.3698;%%

\bibitem{cmssm1} G. Aad et al. (ATLAS collaboration), Phys. Lett. B710 (2012) 67 (2011);
                 Phys. Rev. D 87 (2013) 012008.

\bibitem{cmssm2} S. Chatrchyan et al. (CMS collaboration), Phys. Rev. Lett. 107 (2011) 221804;
                 JHEP 1210 (2012) 018.

\bibitem{han} C.~Han {\it et al.},
  %``Current experimental bounds on stop mass in natural SUSY,''
  JHEP {\bf 1310} (2013) 216
  [arXiv:1308.5307 [hep-ph]].
  %%CITATION = ARXIV:1308.5307;%%

\bibitem{recent} For some recent efforts, see, e.g.,
 J.~Chakrabortty, A.~Choudhury and S.~Mondal,
  %``Non-universal Gaugino mass models under the lamppost of muon (g-2),''
  arXiv:1503.08703 [hep-ph];
  %%CITATION = ARXIV:1503.08703;%%
 K.~Kowalska, L.~Roszkowski, E.~M.~Sessolo and A.~J.~Williams,
  %``GUT-inspired SUSY and the muon g-2 anomaly: prospects for LHC 14 TeV,''
  arXiv:1503.08219 [hep-ph];
  %%CITATION = ARXIV:1503.08219;%%
 K.~Harigaya, T.~T.~Yanagida and N.~Yokozaki,
  %``Higgs mass 125 GeV and g-2 of the muon in Gaugino Mediation Model,''
  arXiv:1501.07447 [hep-ph];
  %%CITATION = ARXIV:1501.07447;%%
 M.~A.~Ajaib, I.~Gogoladze and Q.~Shafi,
  %``GUT-Inspired Supersymmetric Model for h\rightarrow \gamma \gamma and Muon g-2,''
  arXiv:1501.04125 [hep-ph];
  %%CITATION = ARXIV:1501.04125;%%
 F.~F.~Deppisch, N.~Desai and T.~E.~Gonzalo,
  %``Compressed and Split Spectra in Minimal SUSY SO(10),''
  Front.\ Phys.\  {\bf 2}, 00027 (2014)
  [arXiv:1403.2312 [hep-ph]].
  %%CITATION = ARXIV:1403.2312;%%

\bibitem{ewsusy}]
T. Cheng, J. Li, T. Li, D. V. Nanopoulos and C. Tong, Eur. Phys. J. C 73, 2322 (2013);
T. Li, S. Raza,Phys. Rev. D 91, 055016 (2015)

\bibitem{gsugra} S. Akula, P. Nath, Phys. Rev. D 87, 115022 (2013).

\bibitem{ngaugino} Nidal Chamoun, Chao-Shang Huang, Chun Liu, Xiao-Hong Wu, Nucl.\ Phys.\ B624 (2002) 81-94;

  Katri Huitu, Jari Laamanen, Pran N. Pandita, Sourov Roy,Phys.Rev. D72 (2005) 055013;
  
  Katri Huitu, Ritva Kinnunen, Jari Laamanen, Sami Lehti, Sourov Roy, Tapio Salminen, Eur.\ Phys.\ J.\ C58:591-608(2008);
  
  Joydeep Chakrabortty, Amitava Raychaudhuri, Phys.\ Lett.\ B 673:57-62(2009);
  
  Stephen P. Martin, Phys.\ Rev.\ D79:095019(2009);
  
  D. Horton, G. G. Ross,	Nucl.\ Phys.\ B830:221-247(2010);
  
   Frank F. Deppisch, Nishita Desai, Tomas E. Gonzalo, Front.\ Physics 2, 00027 (2014).
   
\bibitem{GSUGRA2} C. Balazs, T. Li, D. V. Nanopoulos, F. Wang, JHEP 1009:003,2010;
T. Li and D. V. Nanopoulos, Phys. Lett. B 692, 121 (2010);
C. Balazs, T. Li, D. V. Nanopoulos, F. Wang, JHEP 1102:096,2011;
F. Wang, Nucl.Phys.B851:104-142,2011.

\bibitem{mSUGRA}
A.~H.~Chamseddine, R.~L.~Arnowitt and P.~Nath,
%``Locally Supersymmetric Grand Unification,''
Phys.\ Rev.\ Lett.\ {\bf 49}, 970 (1982);
%%CITATION = PRLTA,49,970;%%
H.~P.~Nilles,
%``Dynamically Broken Supergravity And The Hierarchy Problem,''
Phys.\ Lett.\ B {\bf 115}, 193 (1982);
%%CITATION = PHLTA,B115,193;%%
L.~E.~Ibanez,
%``Locally Supersymmetric SU(5) Grand Unification,''
Phys.\ Lett.\ B {\bf 118}, 73 (1982);
%%CITATION = PHLTA,B118,73;%%
R.~Barbieri, S.~Ferrara and C.~A.~Savoy,
%``Gauge Models With Spontaneously Broken Local Supersymmetry,''
Phys.\ Lett.\ B {\bf 119}, 343 (1982);
%%CITATION = PHLTA,B119,343;%%
H.~P.~Nilles, M.~Srednicki and D.~Wyler,
%``Weak Interaction Breakdown Induced By Supergravity,''
Phys.\ Lett.\ B {\bf 120}, 346 (1983);
%%CITATION = PHLTA,B120,346;%%
J.~R.~Ellis, D.~V.~Nanopoulos and K.~Tamvakis,
%``Grand Unification In Simple Supergravity,''
Phys.\ Lett.\ B {\bf 121}, 123 (1983);
%%CITATION = PHLTA,B121,123;%%
J.~R.~Ellis, J.~S.~Hagelin, D.~V.~Nanopoulos and K.~Tamvakis,
%``Weak Symmetry Breaking By Radiative Corrections In Broken Supergravity,''
Phys.\ Lett.\ B {\bf 125}, 275 (1983);
%%CITATION = PHLTA,B125,275;%%
N. Ohta,
%%%``Grand Unified Theories Based on Local Supersymmetry,''
Prog.\ Theor.\ Phys.\ 70 (1983) 542;
L.~J.~Hall, J.~D.~Lykken and S.~Weinberg,
%``Supergravity As The Messenger Of Supersymmetry Breaking,''
Phys.\ Rev.\ D {\bf 27}, 2359 (1983).
%%CITATION = PHRVA,D27,2359;%%





\bibitem{gaugemediation}
M.~Dine, W.~Fischler and M.~Srednicki,
%``Supersymmetric Technicolor,''
Nucl.\ Phys.\ B {\bf 189}, 575 (1981);
S.~Dimopoulos and S.~Raby,
%``Supercolor,''
Nucl.\ Phys.\ B {\bf 192}, 353 (1981);
M.~Dine and W.~Fischler, Phys.\ Lett.\ B {\bf 110}, 227 (1982);
M. Dine and A. E. Nelson, Phys. Rev. {\bf D48}, 1277 (1993);
M. Dine, A. E. Nelson and Y. Shirman, Phys. Rev. {\bf D51}, 1362 (1995);
M. Dine, A. E. Nelson, Y. Nir and Y. Shirman, Phys. Rev. {\bf D53}, 2658 (1996);
G. F. Giudice and R. Rattazzi, Phys. Rept. {\bf 322}, 419 (1999).

\bibitem{anomalymediation}
L.~Randall and R.~Sundrum,
%``Out of this world supersymmetry breaking,''
Nucl.\ Phys.\ B {\bf 557}, 79 (1999);
% [arXiv:hep-th/9810155];
%%CITATION = HEP-TH 9810155;%%
G.~F.~Giudice, M.~A.~Luty, H.~Murayama and R.~Rattazzi,
%``Gaugino mass without singlets,''
JHEP {\bf 9812}, 027 (1998).
% [arXiv:hep-ph/9810442]

\bibitem{higgssugra} S. Akula, B. Altunkaynak, D. Feldman et al., Phys. Rev. D 85 (2012) 075001;
                   S. Akula, P. Nath, and G. Peim, Phys. Lett. B 717 (2012) 188.

\bibitem{ngaugino1} J. R. Ellis, K. Enqvist, D. V. Nanopoulos, K. Tamvakis, Phys. Lett. B155 (1985) 381.

\bibitem{ngaugino2} M. Drees, Phys. Lett. B158 (1985) 409.

\bibitem{suspect} A. Djouadi, J. Kneur, G. Moultaka, Comput. Phys. Commun. 176, 426 (2007).

\bibitem{darksusy} P. Gondolo {\it et al.}, JCAP 07 (2004) 008.
The code is available from http://www.physto.se/~edsjo/darksusy.

\bibitem{FWY}  F.~Wang, W.~Wang and J.~M.~Yang,
  %``Split supersymmetry under GUT and current dark matter constraints,''
  Eur.\ Phys.\ J.\ C {\bf 74}, 3121 (2014)
  [arXiv:1310.1750 [hep-ph]].
  %%CITATION = ARXIV:1310.1750;%%

\bibitem{bernal} N. Bernal, A. Djouadi, P. Slavich, JHEP0707, 016 (2007);
                 N.~Bernal, JCAP {\bf 0908},022 (2009).

\bibitem{FWY1} F.~Wang, W.~Wang and J.~M.~Yang,
  %``A split SUSY model from SUSY GUT,''
  JHEP {\bf 1503}, 050 (2015)
  [arXiv:1501.02906 [hep-ph]].
  %%CITATION = ARXIV:1501.02906;%%

\bibitem{planck} http://www.sciops.esa.int/SA/PLANCK/docs/Planck 2013 results 16.pdf.

\bibitem{wmap} J. Dunkley et al.[WMAP Collaboration], Astrophys. J. Suppl. 180, 306 (2009).

\bibitem{oblique} G. Altarelli and R. Barbieri, Phys. Lett. B 253, 161 (1991);
M. E. Peskin, T. Takeuchi, Phys. Rev. D 46, 381 (1992).

\bibitem{stuconstraints} LEP and SLD Collaborations, Phys. Rept. 427 (2006) 257.

\bibitem{xenon} E. Aprile et al. [XENON100 Collaboration], Phys. Rev. Lett. 109, 181301 (2012).

\bibitem{lux} D.S. Akerib et al. [LUX Collaboration], arXiv:1310.8214 [astro-ph.CO].

\bibitem{Djoudi}   A.~Djouadi and M.~Drees, \PLB484, 183 (2000);
    G.~Belanger {\it et al.}, Comput.\ Phys.\ Commun.\  {\bf 180}, 747 (2009).
  %%CITATION = CPHCB,180,747;%%
  %%CITATION = PHLTA,B484,183;%%

\bibitem{Carena} M.~S.~Carena {\it et al.}, \NPB577, 88 (2000).
  %%CITATION = NUPHA,B577,88;%%

\bibitem{Hisano:2010ct}
  J.~Hisano, K.~Ishiwata and N.~Nagata, arXiv:1007.2601 [hep-ph].
  %%CITATION = ARXIV:1007.2601;%%

\bibitem{lattice}   H.~Ohki {\it et al.}, \PRD78, 054502 (2008);
    D.~Toussaint and W.~Freeman, \PRL103, 122002 (2009);
    J.~Giedt, A.~W.~Thomas and R.~D.~Young, \PRL103, 201802 (2009).
  %%CITATION = PRLTA,103,201802;%%
  %%CITATION = PRLTA,103,122002;%%
  %%CITATION = PHRVA,D78,054502;%%

\bibitem{moroi} T. Moroi, Phys. Rev. D 53 (1996) 6565 [Erratum ibid. D 56 (1997) 4424].

\bibitem{endo} M. Endo, K. Hamaguchi, S. Iwamoto1 and T. Yoshinaga, JHEP 01(2014)123;

  M.~Badziak, Z.~Lalak, M.~Lewicki, M.~Olechowski and S.~Pokorski, JHEP {\bf 1503} (2015) 003;
  %``Upper bounds on sparticle masses from muon g ? 2 and the Higgs mass and the complementarity of future colliders,''
  H.~Fargnoli, C.~Gnendiger, S.~Pabehr, D.~Stockinger and H.~Stockinger-Kim, JHEP {\bf 1402} (2014) 070;
   %``Two-loop corrections to the muon magnetic moment from fermion/sfermion loops in the MSSM: detailed results,''
 H.~G.~Fargnoli, C.~Gnendiger, S.~Pabehr, D.~Stoeckinger and H.~Stoeckinger-Kim, Phys.\ Lett.\ B {\bf 726} (2013) 717;
   %``Non-decoupling two-loop corrections to $(g-2)$$_{\mu}$ from fermion/sfermion loops in the MSSM,''
 C.~Gnendiger, D.~Stoeckinger and H.~Stoeckinger-Kim, Phys.\ Rev.\ D {\bf 88} (2013) 053005.
   %``The electroweak contributions to $(g-2)_\mu$ after the Higgs boson mass measurement,''
  
 \bibitem{Farinaldo} Farinaldo S. Queiroz, William Shepherd, Phys. Rev. D 89, 095024 (2014).

\end{thebibliography}
\end{document}